# Photoinduced topological phase transitions in strained black phosphorus

Hang Liu,[1,4] Jia-Tao Sun,[1,4*] Cai Cheng,[1] Feng Liu,[2,3†] and Sheng Meng[1,3,4‡]

[1] *Institute of Physics, Chinese Academy of Sciences, Beijing 100190, P. R. China*
[2] *Department of Materials Science and Engineering, University of Utah, Salt Lake City, Utah 84112, USA*
[3] *Collaborative Innovation Center of Quantum Matter, Beijing 100084, P. R. China*
[4] *University of Chinese Academy of Sciences, Beijing 100049, P. R. China*

## Abstract

Photoinduced topological phase transitions (TPTs) in black phosphorous (BP) under compressive strain were investigated by combining Floquet theory and first-principles calculations. Intriguing photo-dressed electronic states including Floquet-Dirac semimetals and Floquet topological insulators have been identified, which can be feasibly engineered by changing the direction, intensity and frequency of incident laser. Remarkably, tunable TPT from type-I to type-II Floquet-Dirac fermions can be realized by irradiating circularly polarized laser (CPL), with an origin rooted in optical Stark effect. Furthermore, the photoinduced Floquet-Dirac phases is shown to host topological surface states that exhibit nonequilibrium electron transport in a direction locked by the helicity of CPL. This work provides useful guidance for experimental observation of Floquet TPTs, and extends optoelectrionic applications of BP to nonequilibrium regime.

Since the experimental observation of quantum spin Hall insulator [1], topological materials with novel electronic states have attracted significant attention. In particular, topological semimetals including Weyl, Dirac, nodal-ring and nodal-chain semimetals exhibit extraordinary quantum properties such as chiral anomaly [2] and Fermi arc surface states [3-5]. Depending on the tilt of the cone around the nodal point, Weyl or Dirac fermion can be classified as type-I and type-II [6]. Type-I fermion exists in high-energy physics; while type-II fermion does not because it violates relativistic Lorentz invariance. Interestingly, both type-I and type-II fermions were found in condensed matter systems by breaking Lorentz invariance [6-12]. Since type-II Weyl/Dirac points connect hole and electron pockets of the same energy, they have a Fermi surface geometry topologically distinct from that of type-I fermion. Consequently, topological phase transitions (TPTs) between type-I and type-II fermions in a single material are very intriguing of high interest, but extremely challenging to achieve in equilibrium [13].

Coherent laser-matter interactions may provide a feasible solution to overcome the difficulties of TPTs in equilibrium conditions. In analogy to Bloch theorem for crystals, photon-dressed electronic states in periodically driven time-domain (known as Floquet states) can be described by Floquet theorem [14]. Based on effective model Hamiltonian, a number of thermally inaccessible Floquet phases and phase transitions have been proposed recently [15-29]. Topologically trivial electronic states can be driven to Floquet topological insulators [15-19,21-24,26-28,30-32], where steady topological states possess robust nonequilibrium edge states similar to that of topological insulators in equilibrium [1,16]. Besides, Floquet Chern insulators [20,25], Floquet Majorana fermions [29] and Floquet-Weyl semimetal [18,19] can also be induced in laser-driven model systems. However, to date, laser-driven Floquet topological states have rarely been explored for real material realizations [31-34]. Especially, tunable Floquet TPTs in a specific material have not been reported.

Here, we predict that a variety of Floquet topological phases can be created in strained black phosphorous (BP) driven by a time-periodic, space-homogeneous and circularly polarized laser (CPL). Most remarkably, tunable transitions between type-I and type-II Floquet-Dirac fermions (FDFs) can be achieved by changing the direction, intensity, and frequency of incident CPL, originated from optical Stark effect. Furthermore, the electron transport direction of surface states is shown to be locked with the helicity of CPL, enabling feasible control of quantum transport by optical means. We believe that laser-engineered BP is a promising platform for experimental realization of novel FDFs, which extends its fascinating functionalities for optoelectronic applications to nonequilibrium regime.

BP is an elemental layered material with AB stacking, whose structure is shown in Fig. 1(a). It is a semiconductor with a narrow direct band gap of 0.33 ± 0.02 eV [35]. The minimal band gap locates at the time-reversal invariant momentum Z (Fig. 1(b)). The top of valence band and the bottom of conduction band possess an even (+) and odd (−) parity, respectively. Upon applying a 2% uniaxial compressive strain along the armchair direction, the direct band gap decreases to zero resulting in an anisotropic Dirac semimetal. Further compression leads to crossing of valence and conduction bands and inverted band parities (Fig. 1(c)). The inversion leads to formation of type-I Dirac nodal ring lying in the Γ-Z-W plane (Fig. 1(d)), which has been observed in experiments [36-38].

To study coherent interactions between the laser and strained BP, we adopt an anticlockwise CPL with a time-dependent vector potential $A(t) = A_0(\cos(\omega t), \sin(\omega t), 0)$ (see Supporting Information for calculation details). The time-periodic and space-homogeneous CPL propagates along the stacking direction (−z) of BP (Fig. 2(a)). The photon energy and amplitude of the CPL are set as $\hbar\omega$ = 0.5 eV and $A_0$ = 150 V/$c$ (corresponding to 0.038 V/Å or $1.9 \times 10^{10}$ W/cm$^2$, here $c$ is velocity of light), respectively. We find that the Dirac nodal ring in equilibrium is simultaneously driven to a pair of type-I Floquet-Dirac nodal points along the Γ-Z-Γ path, and topologically nontrivial gaps emerge on other paths in the Γ-Z-W plane, as shown in Fig. 2(c,d). Interestingly, when the amplitude of CPL increases to $A_0$ = 300 V/$c$, type-I nodal points are taken over by type-II nodal points (see Fig. 2(e,f)). At the same time, the separation between the pair of nodal points decreases from 0.142 Å$^{-1}$ to 0.068 Å$^{-1}$. As the laser amplitude increases further to 325 V/$c$, the two nodal points merge together, forming an anisotropic Floquet semi-Dirac state, with linear and nonlinear dispersions along zigzag ($k_y$) and stacking ($k_z$) directions, respectively (Fig. S2).

Evolution of Fermi surface contours for irradiated BP is shown in Fig. 2(g-i). When the Fermi level is at the energy of type-II nodal points $\varepsilon_F = \varepsilon_D$, the electron and hole pockets touch at the two type-II nodal points (Fig. 2(h)). This is quite different from point-like Fermi surface of type-I nodal point induced by CPL with weak amplitude. When $\varepsilon_F$ is at the energy of $\varepsilon_D \pm$ 1.5 meV, the electron and hole pockets stay away from each other. The special Fermi surface of type-II FDFs guarantees the emergence of extraordinary transport properties [6,39].

The TPT shown above depends not only on laser amplitude, but also on the incident direction of CPL. For example, if the incident direction of CPL is along the zigzag (*y*) direction, we can obtain type-I nodal points, semi-Dirac nodal point, and topologically trivial bandgap in turn with increasing laser amplitude, while type-II nodal points are absent (Fig. S3). If the incident direction of CPL is along the armchair (*x*) direction, Floquet topological insulator, semi-Dirac semimetal, and trivial insulator can be induced successively as laser amplitude increases (Fig. S4). This angular dependence of Floquet states originates from anisotropic atomic structure of BP, suggesting that topological Floquet states can be easily engineered by tuning the incident direction and amplitude of laser.

To study the continuous evolution of TPTs, the phase diagram as functions of laser intensity and incident direction is constructed, as shown in Fig. 3(c). The angle-resolved and time-dependent vector potential $A(t) = A_0(\cos(\omega t), \sin(\omega t)\sin(\theta), \sin(\omega t)\cos(\theta))$ with fixed photon energy $\hbar\omega = 0.5$ eV is used, where $\theta$ is the angle between the propagation direction of CPL (red arrow) and zigzag (*y*) direction (Fig. 3(a)). Emergence of type-I FDFs has a weak dependence on the incident direction. In contrast, type-II FDFs have a strong dependence on the incident direction, and can be obtained with moderate laser intensity. Interestingly, when the propagation direction of CPL deviates from high symmetry path ($\theta \neq 0°, 90°$), nodal points do not appear along the laser propagation direction, but develop an orientation mismatch ($\theta' \neq \theta$) (see Fig. S5 for details). Here $\theta'$ represents the angle of the line connecting two laser-induced nodal points with respect to $k_y$ axis (or Z-W path).

Besides the incident direction and amplitude of CPL, the photon energy of CPL is another degree of freedom to engineer the TPT from type-I to type-II FDFs. Figure 3(d) shows the phase diagram when the incident direction of CPL is restricted to stacking direction ($\theta$ = 90°). We find that the states with type-I FDFs exist in a large range of photon energy and laser amplitude. In contrast, type-II FDFs can only be induced by the CPL with moderate amplitude and infrared photon energy. Consequently, the following conditions to realize TPT from type-I to type-II FDFs are required simultaneously: i) Light propagation is restricted along stacking direction ($\theta$ = 70° ~ 90°); ii) Laser amplitude and photon energy are set in the ranges of $A_0$ = 150 ~ 350 V/$c$ and $\hbar\omega$ = 0.2 ~ 1.0 eV.

Next, to reveal the mechanism of above TPTs, we consider the case of CPL with a vector potential $\boldsymbol{A}(t) = A_0(\cos(\omega t), \sin(\omega t), 0)$ propagating along the stacking direction of BP. Once the CPL is irradiated on BP, the electrons would emit and absorb photons to form photon-dressed (or side) bands labelled by Floquet band index $n$ = ···, −2, −1, 1, 2, ··· (gray thin line in Fig. 4(a)), while the static component is indexed with $n$ = 0 (gray bold line in Fig. 4(a)). The photon-dressed states would hybridize with $n$ = 0 states, known as optical Stark effect, leading to band repulsion $\Delta = \sqrt{A_0^2 |M|^2 + (\delta E)^2} - \delta E$ [34]. Here $M$ is the dipole matrix element between the two states, and $\delta E$ is the energy difference of the two states before hybridization. Because the energy difference between $n$ = 0 and $n$ = −1 bands is zero ($\delta E$ = 0) at their crossing points, the induced gap $\Delta$ increases linearly with laser amplitude $A_0$.

The energy difference between the original nodal point in equilibrium and the crossing point of $n$ = 0 and $n$ = −1 bands on the Γ-Z path is defined as $\Delta'$ (Fig. 4(a)). Obviously, in order to drive the TPT from type-I to type-II FDFs, the optical Stark effect should be strong enough to satisfy $\Delta/2 > \Delta'$. When photon energy is fixed to $\hbar\omega$ = 0.5 eV, the energy difference $\Delta'$ is 0.17 eV (pink dashed line in Fig. 4(c)). The crossing point of the pink dashed line and pink solid line separates type-II (above pink dashed line) from type-I Floquet-Dirac states (below pink dashed line). However, when photon energy increases sufficiently (e.g. $\hbar\omega$ = 1.5 eV), the energy difference $\Delta'$ can be so large that the requirement $\Delta/2 > \Delta'$ cannot be satisfied before the two type-I nodal points merge together (Fig. 4(b)). Consequently, the type-II FDFs can no longer be realized. Furthermore, as shown in Fig. 4(d), the phase boundary between type-I and type-II FDFs can be plotted according to the crossing points of dashed and solid lines in Fig. 4(c). It is seen that the phase boundary results from the linear dependence of $\Delta \sim A_0$ and the linear dispersions around the nodal point. Besides the crucial role of optical Stark effect played in the TPT from type-I to type-II FDFs, the weak dependence of shift of nodal points on photon energy, as shown in Fig. 4(b), is also contributed by this effect, which is missing in previous model Hamiltonian calculations [18].

In equilibrium, the strained BP has a drumhead surface state in the surface Brillouin zone (SBZ) of the (100) surface (Fig. 1(a,b)) [36]. It is natural to ask if nonequilibrium surface states protected from backscattering also appear in the strained BP under light irradiation. When turning on CPL to propagate along the stacking direction ($-z$), the strained BP has the nodal points along Γ-Z-Γ path and topologically nontrivial gaps along other paths (Γ-Z-W plane) (see Fig. 2(d)). Interestingly, the Floquet surface states along $\bar{\Gamma}$-$\bar{Z}$-$\bar{\Gamma}$ path connect two nodal points (Fig. S6), while along other paths on SBZ passing $\bar{Z}$, surface states connect two topologically nontrivial gaps of type-I FDFs. Along $\bar{W}$-$\bar{Z}$-$\bar{W}$ (or $\bar{k}_y$) direction, the surface states of type-I FDFs at two opposite surfaces have opposite slopes (Fig. 5(a-c)). Two Fermi arcs contributed by two opposite surfaces connect two nodal points (Fig. S6).

As sketched in Fig. 5(a-c), if the helicity of CPL is set to be counterclockwise, the transport directions of nonequilibrium surface states on left (right) surface are along $+y$ ($-y$). Once the helicity of CPL is changed to be clockwise, the direction of topologically protected transport channel on each surface would be reversed (Fig. 5(d-f)). The locking effect is independent of the type of FDFs (Fig. S7). Therefore, by changing CPL frequency and amplitude, not only the dispersions of surface states can be tuned, but also the Fermi arc of the FDFs. If the laser incident direction is along other paths in the Γ-Z-W plane, the locking effect of the transport direction with respect to the laser helicity remains valid (Fig. S8). The robust locking for the topological Floquet-Dirac states provides an effective method to control the dissipationless surface states by laser illumination.

In conclusion, a number of nonequilibrium topological phases in the uniaxially compressed BP under the irradiation of CPL have been identified, including various types of FDFs and Floquet topological insulators. The TPTs between them can be engineered by tuning incident direction, intensity and photon energy of the CPL. The intriguing TPT from type-I to type-II FDFs is predicted when infrared laser with moderate strength propagates along the stacking direction of BP. The transport directions of novel nonequilibrium surface states, resulted from bulk-boundary correspondence of the topological Floquet-Dirac states, are locked with the helicity of CPL, providing the possibility to optically control nonequilibrium quantum transport properties.

We gratefully acknowledge financial support from the National Key R&D Program of China (Grant No. 2016YFA0202300, 2016YFA0300902), National Basic Research Program of China (Grant No. 2013CBA01600), and "Strategic Priority Research Program (B)" of Chinese Academy of Sciences (Grant No. XDB07030100). F. L. was supported by US DOE-BES (No. DE-FG02-04ER46148).

[*] jtsun@iphy.ac.cn
[†] fliu@eng.utah.edu
[‡] smeng@iphy.ac.cn

[1] M. Koenig, S. Wiedmann, C. Bruene, A. Roth, H. Buhmann, L. W. Molenkamp, X.-L. Qi, and S.-C. Zhang, Science **318**, 766 (2007).
[2] X. C. Huang, L. X. Zhao, Y. J. Long, P. P. Wang, D. Chen, Z. H. Yang, H. Liang, M. Q. Xue, H. M. Weng, Z. Fang, X. Dai, and G. F. Chen, Phys. Rev. X **5**, 031023 (2015).
[3] X. Wan, A. M. Turner, A. Vishwanath, and S. Y. Savrasov, Phys. Rev. B **83**, 205101 (2011).
[4] S.-Y. Xu *et al.*, Science **349**, 613 (2015).
[5] B. Q. Lv, H. M. Weng, B. B. Fu, X. P. Wang, H. Miao, J. Ma, P. Richard, X. C. Huang, L. X. Zhao, G. F. Chen, Z. Fang, X. Dai, T. Qian, and H. Ding, Phys. Rev. X **5**, 031013 (2015).
[6] A. A. Soluyanov, D. Gresch, Z. Wang, Q. Wu, M. Troyer, X. Dai, and B. A. Bernevig, Nature **527**, 495 (2015).
[7] Z. K. Liu, B. Zhou, Y. Zhang, Z. J. Wang, H. M. Weng, D. Prabhakaran, S.-K. Mo, Z. X. Shen, Z. Fang, X. Dai, Z. Hussain, and Y. L. Chen, Science **343**, 864 (2014).
[8] H. M. Weng, C. Fang, Z. Fang, B. A. Bernevig, and X. Dai, Phys. Rev. X **5**, 10, 011029 (2015).
[9] B. Q. Lv, N. Xu, H. M. Weng, J. Z. Ma, P. Richard, X. C. Huang, L. X. Zhao, G. F. Chen, C. E. Matt, F. Bisti, V. N. Strocov, J. Mesot, Z. Fang, X. Dai, T. Qian, M. Shi, and H. Ding, Nat. Phys. **11**, 724 (2015).
[10] H. Huang, S. Zhou, and W. Duan, Phys. Rev. B **94**, 121117 (2016).
[11] L. Huang, T. M. McCormick, M. Ochi, Z. Zhao, M. T. Suzuki, R. Arita, Y. Wu, D. Mou, H. Cao, J. Yan, N. Trivedi, and A. Kaminski, Nat. Mater. **15**, 1155 (2016).
[12] Z. Wang, Y. Sun, X.-Q. Chen, C. Franchini, G. Xu, H. Weng, X. Dai, and Z. Fang, Phys. Rev. B **85**, 195320 (2012).
[13] G. E. Volovik, Low Temp. Phys. **43**, 47 (2017).
[14] A. Gomez-Leon and G. Platero, Phys. Rev. Lett. **110**, 200403 (2013).

[15] A. Farrell and T. Pereg-Barnea, Phys. Rev. B **93**, 045121 (2016).

[16] N. H. Lindner, G. Refael, and V. Galitski, Nat. Phys. **7**, 490 (2011).

[17] T. Oka and H. Aoki, Phys. Rev. B **79**, 081406 (2009).

[18] Z. Yan and Z. Wang, Phys. Rev. Lett. **117**, 087402 (2016).

[19] C.-K. Chan, Y.-T. Oh, J. H. Han, and P. A. Lee, Phys. Rev. B **94**, 121106(R) (2016).

[20] A. G. Grushin, A. Gomez-Leon, and T. Neupert, Phys. Rev. Lett. **112**, 156801 (2014).

[21] T. Kitagawa, T. Oka, A. Brataas, L. Fu, and E. Demler, Phys. Rev. B **84**, 235108 (2011).

[22] Y. T. Katan and D. Podolsky, Phys. Rev. Lett. **110**, 016802 (2013).

[23] G. Usaj, P. M. Perez-Piskunow, L. E. F. Foa Torres, and C. A. Balseiro, Phys. Rev. B **90**, 115423 (2014).

[24] M. A. Sentef, M. Claassen, A. F. Kemper, B. Moritz, T. Oka, J. K. Freericks, and T. P. Devereaux, Nat. Commun. **6**, 7047 (2015).

[25] L. D'Alessio and M. Rigol, Nat. Commun. **6**, 8336 (2015).

[26] M. Ezawa, Phys. Rev. Lett. **110**, 026603 (2013).

[27] P. M. Perez-Piskunow, G. Usaj, C. A. Balseiro, and L. E. F. F. Torres, Phys. Rev. B **89**, 121401 (2014).

[28] J. Inoue and A. Tanaka, Phys. Rev. Lett. **105**, 017401 (2010).

[29] A. A. Reynoso and D. Frustaglia, Phys. Rev. B **87**, 115420 (2013).

[30] M. S. Rudner, N. H. Lindner, E. Berg, and M. Levin, Phys. Rev. X **3**, 031005 (2013).

[31] Y. H. Wang, H. Steinberg, P. Jarillo-Herrero, and N. Gedik, Science **342**, 453 (2013).

[32] F. Mahmood, C.-K. Chan, Z. Alpichshev, D. Gardner, Y. Lee, P. A. Lee, and N. Gedik, Nat. Phys. **12**, 306 (2016).

[33] H. Hubener, M. A. Sentef, U. De Giovannini, A. F. Kemper, and A. Rubio, Nat. Commun. **8**, 13940 (2017).

[34] U. De Giovannini, H. Hubener, and A. Rubio, Nano Lett. **16**, 7993 (2016).

[35] M. Batmunkh, M. Bat-Erdene, and J. G. Shapter, Adv. Mater. **28**, 8586 (2016).

[36] J. Zhao, R. Yu, H. Weng, and Z. Fang, Phys. Rev. B **94**, 195104 (2016).

[37] Z. J. Xiang, Phys. Rev. Lett. **115**, 186403 (2015).

[38] C.-H. Li, Y.-J. Long, L.-X. Zhao, L. Shan, Z.-A. Ren, J.-Z. Zhao, H.-M. Weng, X. Dai, Z. Fang, C. Ren, and G.-F. Chen, Phys. Rev. B **95**, 125417 (2017).

[39] Z. M. Yu, Y. Yao, and S. A. Yang, Phys. Rev. Lett. **117**, 077202 (2016).

**Figures:**

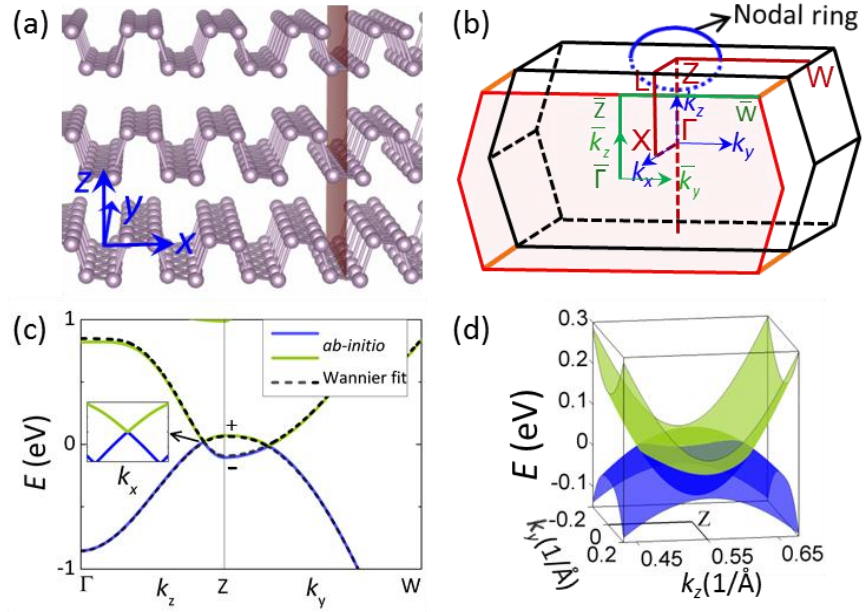

FIG. 1. Band structure of compressive BP in equilibrium. (a) Atomic structure of BP. The *x*, *y* and *z* axes are along armchair, zigzag and stacking directions respectively. The vertical brown plane is the truncated surface used to calculate surface states. (b) Bulk and projected (100) surface first Brillouin zone. The nodal ring (blue line circle) lies in the Γ-Z-W plane. (c) Band structure of BP with 3.72% compression along armchair direction. Solid and dashed lines are the results from *ab-initio* calculations and Wannier fitting respectively. Inset shows band structure along the path that is parallel to $k_x$ direction and crosses a point in the nodal ring along ΓZ direction. (d) Band structure diagram of (c) in the Γ-Z-W plane. The lower band (blue) and upper band (green) cross to form the nodal ring.

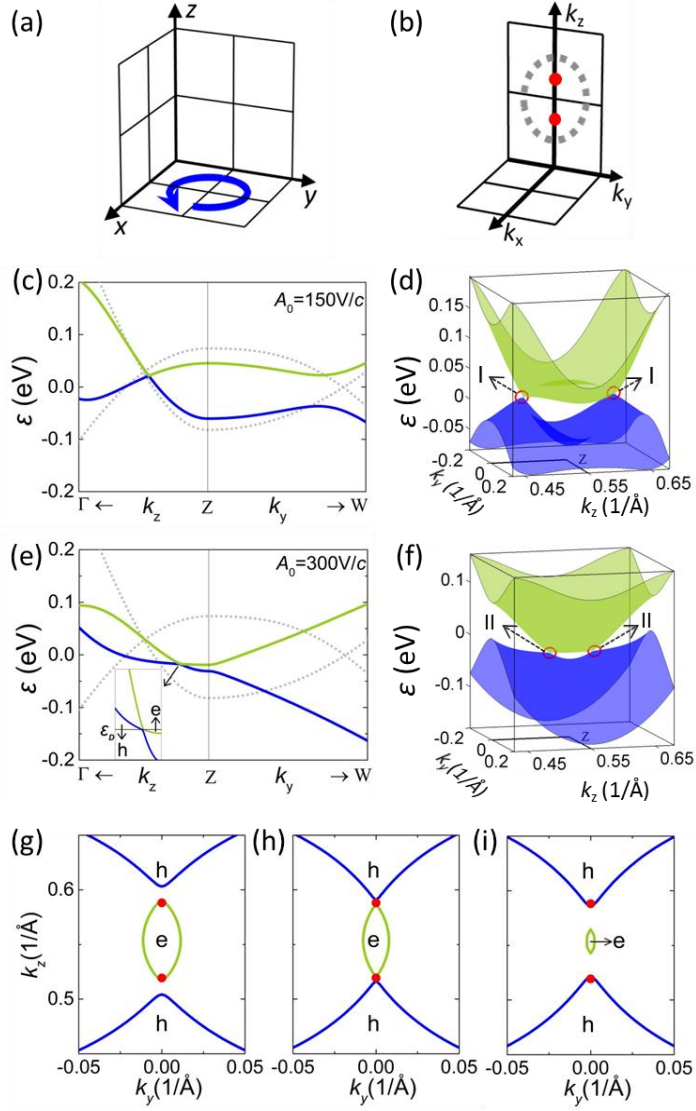

FIG. 2. Topological FDFs induced by laser with photon energy $\hbar\omega$ = 0.5 eV. (a) CPL $A(t) = A_0(\cos(\omega t),\sin(\omega t),0)$ propagates along stacking direction ($-z$). (b) The laser-induced Floquet-Dirac nodal points (red points) and original nodal ring in equilibrium (dashed circle). Floquet-Bloch band structure and band diagram of BP driven by laser with amplitude $A_0$ = 150 V/$c$ (c,d) and $A_0$ = 300 V/$c$ (e,f). Gray dotted line is the equilibrium electronic structure. I and II represent laser-induced type-I and type-II Floquet-Dirac nodal points respectively. The quasienergy of nodal points is marked as $\varepsilon_D$. (g-i) Fermi contour lines on the $k_x$ = 0 ($\Gamma$-Z-W) plane when Fermi energy is at $\varepsilon_D$ + 1.5 meV, $\varepsilon_D$, $\varepsilon_D$ − 1.5 meV respectively. The red dots are the positions of projected type-II nodal points on $k_x$ = 0 plane. The green closed ellipses and blue open hyperbolas represent the contours of the electron and hole pockets in the plane $k_x$ = 0 respectively.

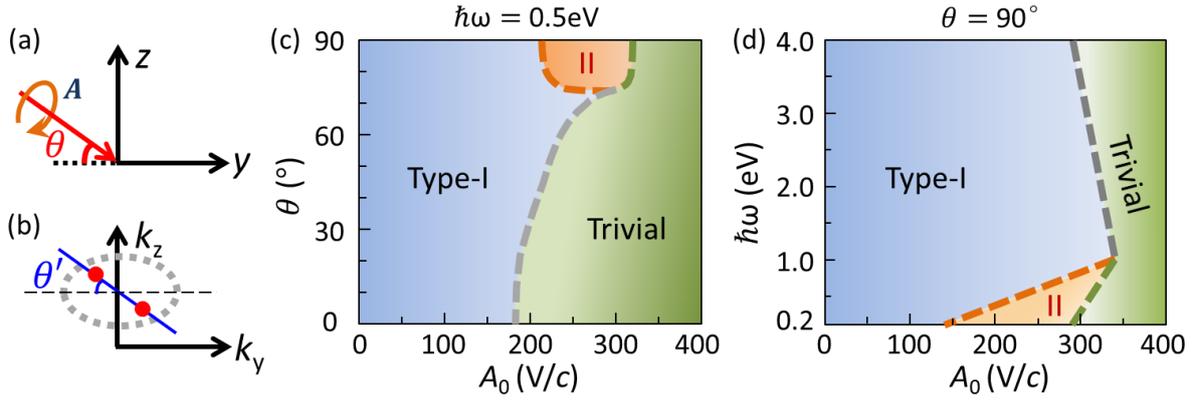

FIG. 3. Laser-induced Floquet phase diagram of compressively strained BP. (a) Anticlockwise CPL propagates on the *yz* plane with the propagation direction as $\theta$. (b) The gray circle represents equilibrium nodal ring of strained BP. The angle between connecting line of two Floquet-Dirac nodal points (red dots) and $k_y$ direction is $\theta'$. (c) Phase diagram of laser-driven BP (photon energy $\hbar\omega = 0.5$ eV) on the dependence of laser amplitude $A_0$ and incident angle $\theta$. (d) The Floquet phases induced by laser ($\theta = 90°$) with different amplitude $A_0$ and frequency $\omega$.

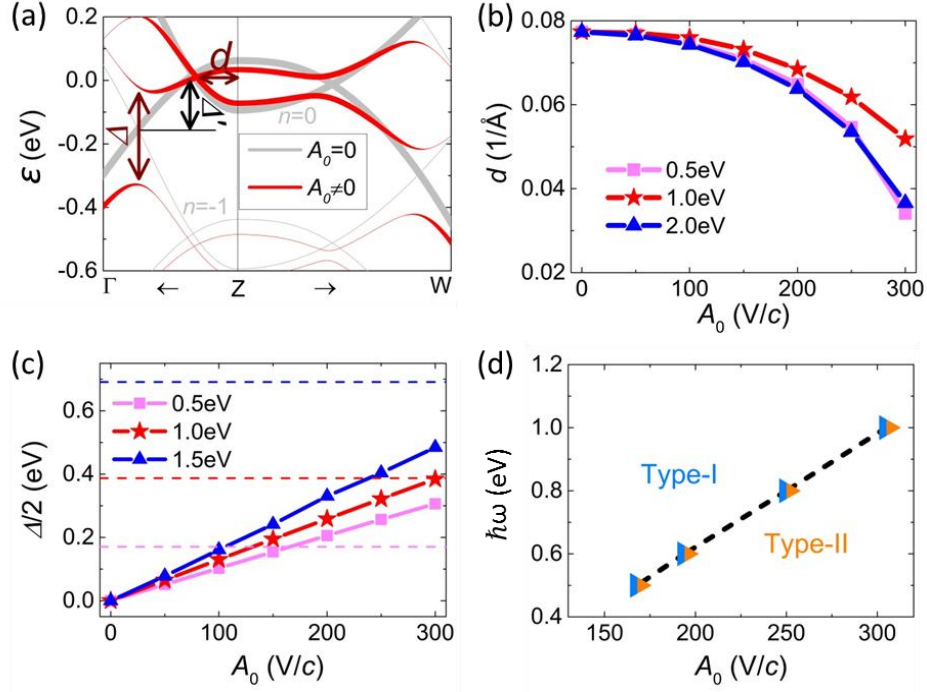

FIG. 4. Origin of the Floquet phase transition of BP driven by CPL propagating along stacking direction (−z). (a) The bold and thin gray lines represent $n = 0$ and $n = -1$ Floquet-Bloch bands respectively. $\Delta'$ is the energy difference between Dirac point on nodal ring and the crossing of $n = 0$ and $n = -1$ bands along Γ-Z direction. The thickness of the line is proportional to the weight of the static ($n = 0$) component ($|\langle u_\alpha^0 | u_\alpha^0 \rangle|^2$). Band gap $\Delta$ is induced by the hybridization between bands indexed by $n = 0$ and $n = -1$. The distance between Floquet-Dirac nodal point and center of the nodal ring is marked as $d$. (b) The variation of $d$ with laser amplitude $A_0$ when the photon energy is set as $\hbar\omega$ = 0.5, 1.0 and 2.0 eV respectively. (c) The dependence of $\Delta/2$ on laser amplitude $A_0$ when photon energy is set as $\hbar\omega$ = 0.5, 1.0 and 1.5 eV respectively. Dashed lines represent the value $\Delta'$ in three cases. (d) Linear dependence of the photon energy and laser amplitude defining the phase boundary of type-I and type-II Floquet-Dirac phases.

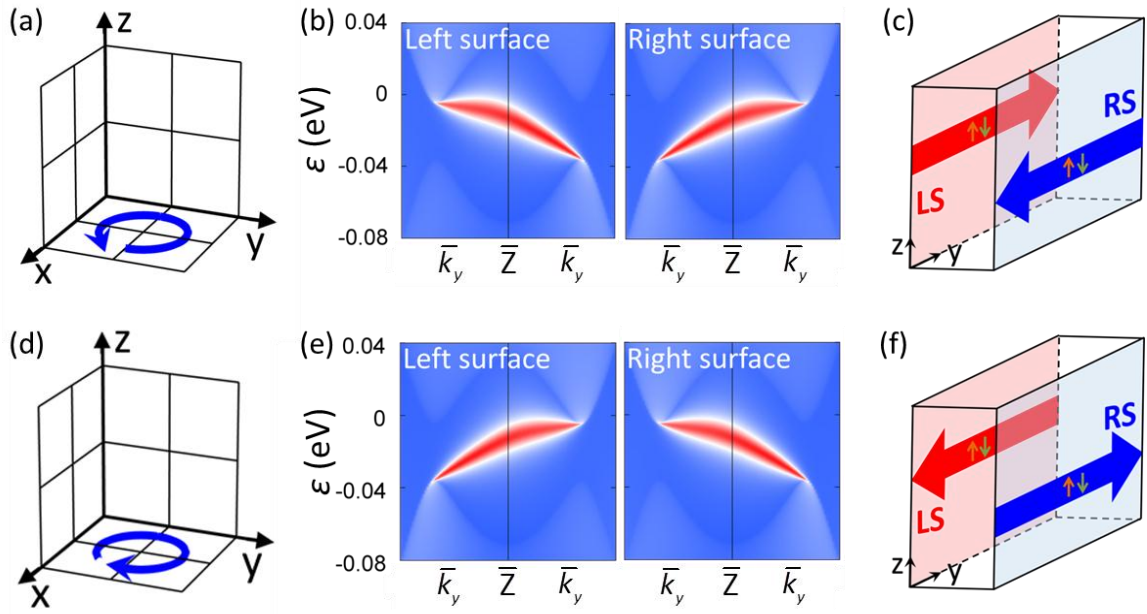

FIG. 5. The locking effect of the transport direction of surface states and laser helicity. (a) The sketch of laser helicity. (b) Surface states along $\bar{W}$-$\bar{Z}$-$\bar{W}$ direction in SBZ come from left and right surfaces respectively, which is induced by laser with photon energy $\hbar\omega$ = 0.9 eV and amplitude $A_0$ = 150 V/c. (c) The opposite direction of topologically protected surface currents on two counter surfaces. LS: left surface; RS: right surface. (d-f) Surface states of Floquet-Dirac state when the laser helicity is reversed.